\documentclass{ws-procs975x65}

\begin{document}

\title{\uppercase{Primordial non-Gaussianity as a signature of pre-inflationary
  radiation era}}

\author{\uppercase{Suratna Das$^*$}}

\address{Tata Institute of Fundamental Research, Mumbai 400005, India\\
$^*$E-mail: suratna@tifr.res.in}

\begin{abstract}
Primordial non-Gaussianity generated in an inflationary model where
inflation is preceded by a radiation era is discussed. It is shown
that both bispectrum and trispectrum non-Gaussianities are enhanced
due to the presence of pre-inflationary radiation era. One
distinguishing feature of such a scenario is that the trispectrum
non-Gaussianity is larger than the bispectrum one.
\end{abstract}


\bodymatter

\section{Introduction}

In a scenario where inflation is preceded by a radiation era, the
inflaton field can initially be in thermal equilibrium with the
radiation field before decoupling and thus can retain its thermal
distribution at the onset of inflation \cite{Bhattacharya:2005wn}. Due
to this initial thermal distribution of inflaton field the power
spectrum of the primordial perturbations is modified by an extra
temperature dependent factor ($\coth(k/2T)$) which enhances the power
of the $TT$ angular power spectrum of CMBR at largest scales. This
angular power spectrum of CMBR is in accordance with the observations
of WMAP if the comoving temperature $T$ of the primordial
perturbations is less than $10^{-3}$ Mpc$^{-1}$ and thus cannot
discriminate this scenario with a generic super-cool inflationary
scenario at the level of power spectrum.

On the other hand, primordial non-Gaussianity (NG)
\cite{Bartolo:2005ae}, which is a measure of departure from a Gaussian
distribution of primordial fluctuations generated during inflation,
has become very crucial for both theoretical and observational
cosmology as measuring or simply constraining NG can discriminate
between several degenerate inflationary models which produce the same
scale-invariant power spectrum and thus can quantify the dynamics of
very early universe. The non-Gaussian parameter $f_{NL}$ arising from
three-point correlation function, known as Bispectrum, is constrained
by WMAP 5yr data \cite{Komatsu:2008hk} as $-151<f_{NL}<253$ whereas
PLANCK \cite{Komatsu:2001rj} can probe up to $f_{NL}\sim 5$. Another
non-Gaussian parameter $g_{NL}$ arising from four-point correlation
function, known as Trispectrum, is loosely constrained by WMAP
\cite{Boubekeur:2005fj} as $|\tau_{NL}|<10^8$ whereas PLANCK
\cite{Kogo:2006kh} can probe $\tau_{NL}$ up to 560. But measurement of
primordial non-Gaussianities imprinted in 21 cm background radiation
in future experiments can constrain the bispectrum
\cite{Cooray:2006km} as $f_{NL}<1$ and the trispectrum
\cite{Cooray:2008eb} as $\tau_{NL}\sim10$.

We will discuss here the non-Gaussian features of such a scenario
\cite{Das:2009sg} where inflation is preceded by a radiation era and
will show that at the level of trispectrum this scenario differs from
a super-cool inflation scenario which can be considered as a signature
of pre-inflationary radiation era.

\section{Enhancement of Bispectrum}

In a generic inflationary scenario a non-Gaussian feature arises due
to non-linear evolution of comoving curvature perturbation
$\mathcal{R}(t,\mathbf x)$ as 
\begin{eqnarray}
{\mathcal R}_{NL}(t,{\mathbf x})=\frac{H}{\dot{\phi}}\delta\phi_L(t,{\mathbf x})+\frac12\frac{\partial}{\partial\phi}\left(\frac{H}{\dot{\phi}}\right)\delta\phi_L^2(t,{\mathbf x})+{\mathcal O}(\delta\phi_L^3),
\end{eqnarray}
though the inflaton fluctuations $\delta\phi_L(t,{\mathbf x})$ are
Gaussian in nature. This kind of non-linear evolution of comoving
curvature perturbations yields a non-vanishing three-point correlation
function or the Bispectrum which can be quantified as 
\begin{eqnarray}
\left\langle{\mathcal R}({\mathbf k}_1){\mathcal R}({\mathbf k}_2){\mathcal R}({\mathbf k}_3)\right\rangle=\frac{\delta^3({\mathbf k_1}+{\mathbf k_2}+{\mathbf k_3})}{(2\pi)^{\frac32}}\frac65 f_{NL}\left(\frac{P_{\cal R}(k_1)}{k_1^3}\frac{P_{\cal R}(k_2)}{k_2^3}+2\,\,{\rm perms}\right),
\end{eqnarray}
where the non-linear parameter in a generic inflationary scenario
turns out to be of the order of slow-roll parameters
($f_{NL}=\frac56(\delta-\epsilon)$) \cite{Byrnes:2006vq} and thus too
small ($\mathcal{O}(10^{-2})$) to be detected by any present or
forthcoming experiments. $P_{\cal R}$ in the above equation represents
the two-point correlation function, called the power spectrum, and the
delta function ensures that the three momentum form a triangle due to
which $f_{NL}$ is quantified in several triangle configurations namely
(i) {\it Squeezed configuration} $(|\mathbf{k}_1|\approx
|\mathbf{k}_2|\approx k\gg |\mathbf{k}_3|)$, (ii) {\it Equilateral
  configuration} $(|\mathbf{k}_1|=|\mathbf{k}_2|=|\mathbf{k}_3|=k)$ and
(iii) {\it Folded configuration}
$(|\mathbf{k}_1|=|\mathbf{k}_2|=\frac12|\mathbf{k}_3|=k)$.

If inflation is preceded by a radiation era then 
\vspace{-0.2cm}
\begin{itemize}
\item the inflaton field will have an initial thermal distribution
\item the thermal vacuum $|\Omega\rangle$ will have finite occupation
  $N_k|\Omega\rangle=n_k|\Omega\rangle$
\item the probability of the system to be found in an energy state
  $\epsilon_r$ would be 
$p(k_1,k_2, k_3,k_4)\equiv\frac{\prod_re^{-\beta n_{k_r}
k_r}}{\prod_r\sum_{n_k}e^{-\beta n_{k_r}
k_r}}=\frac{\prod_re^{-\beta n_{k_r} k_r}}{Z}$.
\end{itemize}
\vspace{-0.2cm}
Due to these above mentioned points the correlation functions
(two-point, three-point and four-point) now have to be thermally
averaged. The thermally averaged two-point correlation function yields
the extra $\coth(k/2T)$ factor \cite{Bhattacharya:2005wn} in the power
spectrum as has been mentioned before. Thermal averaging of the
three-point correlation function enhances the bispectrum NG
\cite{Das:2009sg} . The enhanced $f_{NL}$ has been computed in
different triangle configuration and it is most enhanced in the
Equilateral configuration where it is enhanced by a factor of 91 with
respected to that of in generic inflationary scenarios. Thus it would
be in the range of observation of future experiments of 21 cm
background radiation \cite{Cooray:2006km} .
\vspace{-0.3cm}
\section{Enhancement of Trispectrum}

The trispectrum is known as the connected part of the four-point
correlation function of primordial perturbation and can be written as
\begin{eqnarray}
\left\langle{\mathcal R}({\mathbf k}_1){\mathcal R}({\mathbf k}_2){\mathcal R}({\mathbf k}_3){\mathcal R}({\mathbf k}_4)\right\rangle_c&\equiv&\left\langle{\mathcal R}({\mathbf k}_1){\mathcal R}({\mathbf k}_2){\mathcal R}({\mathbf k}_3){\mathcal R}({\mathbf k}_4)\right\rangle\nonumber\\
&-&\left(\left\langle{\mathcal R}_{L}({\mathbf k}_1){\mathcal R}_{L}({\mathbf k}_2)\right\rangle\left\langle{\mathcal R}_{L}({\mathbf k}_3){\mathcal R}_{L}({\mathbf k}_4)\right\rangle+2\,\,{\rm perm}\right).
\end{eqnarray}
In a generic inflationary scenario the non-Gaussianity arising from
trispectrum, quantified by the parameter $\tau_{NL}$, turns out to be
$\tau_{NL}=\left(\frac65 f_{NL}\right)^2$ \cite{Byrnes:2006vq}
i.e. $\mathcal{O}\left(10^{-4}\right)$. Thus a generic scenario of
inflation yields a trispectrum non-Gaussianity which is much smaller
than a bispectrum non-Gaussianity.

In a presence of pre-inflationary radiation era one has to thermal
average over the four-point correlation function as has been discussed
before. It turns out that after thermal averaging the four-point
correlation function is not equal to the square of thermal average of
two-point correlation function. Thus a non-vanishing four-point
correlation function exists even without non-linear terms in the
comoving curvature perturbations. This shows that the $|\tau_{NL}|$
will not depend upon the slow-roll parameters and can be as large as
42 \cite{Das:2009sg} and which is withing the detection range of
future 21 cm background radiation experiments \cite{Cooray:2008eb}.

\section{Conclusion}

Presence of pre-inflationary radiation era yields an initial thermal
distribution of the inflaton fluctuations. If the initial temperature
of these fluctuations is less than $10^{-3}$ Mpc$^{-1}$ then the power
spectrum is in accordance with the observation and does remain
indistinguishable from the generic case at the level of power
spectrum. On the other hand, such a pre-inflationary radiation era
enhances primordial non-Gaussianity in both bispectrum and trispectrum
level and these non-Gaussianities are within the observable range of
the future experiments of primordial non-Gaussianities imprinted in
the 21 cm background radiation. Interestingly we found out the
trispectrum in this case does not depend upon slow-roll parameters due
to thermal averaging and is orders of magnitude larger than the
bispectrum non-Gaussianity. Thus this very feature of primordial
non-Gaussianities can be considered as a signature of pre-inflationary
radiation era.

\section*{Acknowledgements}

I would like to thank my collaborator of this work Subhendra Mohanty.

\end{document}